\documentclass[]{spie}  
\usepackage[]{graphicx}
\usepackage{color}

\title{Biased Moments of Undersampled Sources}
\author{Andrew K. Bradshaw
\skiplinehalf
Kavli Institute for Particle Astrophysics and Cosmology, Stanford University \\
SLAC National Accelerator Laboratory, Menlo Park, CA 94025\\
}

\authorinfo{Further author information: Send correspondence to A.K.B.: bradshaw@slac.stanford.edud}

\begin{document} 
\maketitle 

\begin{abstract}
Spatial intensity moments computed on images can be used as a probe of the centroid, size, and orientation of pixelized sources such as stars and galaxies. However, all measurements made on images suffer from errors due to undersampling and finite pixel size, causing systematic biases in the computation of moments and other statistics. We show examples of bias in the first and second geometric moments computed on images of Gaussian sources with widths near the pixel scale, $0.1<\sigma<5$ pixels. We then illustrate how undersampling could lead to the orientation angle patterns seen in various modern surveys of the sky.
\end{abstract}


\section{Introduction}
\label{sec:intro} 
The localization and measurement of light sources, such as estimation of the centroid, ellipticity, and orientation of a galaxy, is often a prime focus of astronomical surveys of the sky. One such set of image measurements is the method of moments which summarizes spatial information about the intensity distribution of an object and is capable of uniquely characterizing a distribution function \cite{Hu1962InfoTheory}. For instance, the geometric moments $m_{pq}$ of a 2D image $I$ are calculated by summing the product of pixel values $I_{i,j}$ with coordinates $(x_i,y_j)$ to the $\{p,q\}$ power: $m_{pq} = \sum_i \sum_j x_i^p y_j^q I_{ij}w_{ij}/ \sum_i \sum_j I_{ij}w_{ij}$, where $w_{ij}$ is a pixel weighting function that can be chosen for the application. The zeroth moment $m_{00}$ represents the total flux (often normalized to unity), and the first moment's estimate of the centroid is $\{\bar{x}, \bar{y} \} = \{m_{10}/m_{00}, m_{01}/m_{00} \}$. Higher moments calculated about that centroid are the central moments, $\mu_{pq} = \sum_i \sum_j (x_i - \bar{x})^p(y_j - \bar{y})^q I_{ij}w_{ij}/ \sum_i \sum_j I_{ij}w_{ij}$, where the second, third, and fourth central moments provide information about the size/shape, skewness, and tails of the object being imaged. A perfectly-sampled elliptical 2D Gaussian is therefore completely summarized by moments up to order two, where the second central moments in $x$ and $y$ equal the Gaussian widths $\sigma_x,\sigma_y = \sqrt{\mu_{2,0}}, \sqrt{\mu_{0,2}}$ and the cross term $\mu_{1,1}$ can be used to measure the orientation angle relative to the measurement axes, $\theta = \frac{1}{2} \arctan (2\mu_{1,1}/(\mu_{2,0} - \mu_{0,2}))$.

With astronomical images these calculations are performed on two-dimensional grids of pixel values which correspond to spatial integrals of an object's light over a pixel area and sampled at discrete positions. However, this necessarily breaks the unique correspondence between intrinsic properties and image moments, which requires the image to be point-sampled at continuous positions rather than integrated over areas at discrete positions (as most realistic imagers do). For instance, the centroid of a half-pixel-wide object will be correct only when located at a pixel center or edge; elsewhere the first moment is biased by the assumption that flux is point-sampled at pixel centers. Unbiased estimation of an object's true size and shape using moments measured on pixelized images therefore requires assumptions about both the pixel response function and the sub-pixel distribution of light. For objects much larger than the pixel scale this bias usually becomes negligible compared to statistical noise, but in general it remains a systematic bias in every measurement made on pixelized images.

We illustrate the effect of this bias on second moment (size) measurement in one dimension using a Gaussian source $g(x)$ of variable width $\sigma$ and centered at $x=0$, noiselessly imaged by a linear array pixels indexed by $x_i$ and spaced by $\Delta_x$. The resultant digital image $g_i$ can be represented as the integral of the continuous object's intensity distribution $g(x)$ over a pixel response function, typically assumed to be a unit square $rect(u)$ which is nonzero only for $|u|<1/2$:
\begin{equation}
    g_{rect} (x_i) =\int g(x) \times rect(\frac{x-x_i}{\Delta_x}) dx = 
    \sigma\sqrt{\frac{\pi}{2}}(erf(x_i-\Delta_x/2) - erf(x_i+\Delta_x/2)) .
    \label{eq:1d_pixel_samp}
\end{equation}
Under these conditions, the pixel values of $g_{rect}$ are differences of error functions which can be calculated through numerical integration and fast and accurate approximations. In the limit of infinitely small pixel widths, the pixel response can be modeled as a Dirac $\delta$-sampled image $g_\delta$ which is simply the Gaussian evaluated at each integer pixel position. The difference between these two images is a bias $b$ which can be approximated to second order via a Taylor series approximation \cite{Hagen2008ApOpt}:

\begin{equation}
    b(x_i)=\frac{x^2}{24\sigma^4} exp(-x_i^2/(2\sigma^2)).
    \label{eq:bias_taylor}
\end{equation}

\begin{figure}\includegraphics[width=1.0\columnwidth]{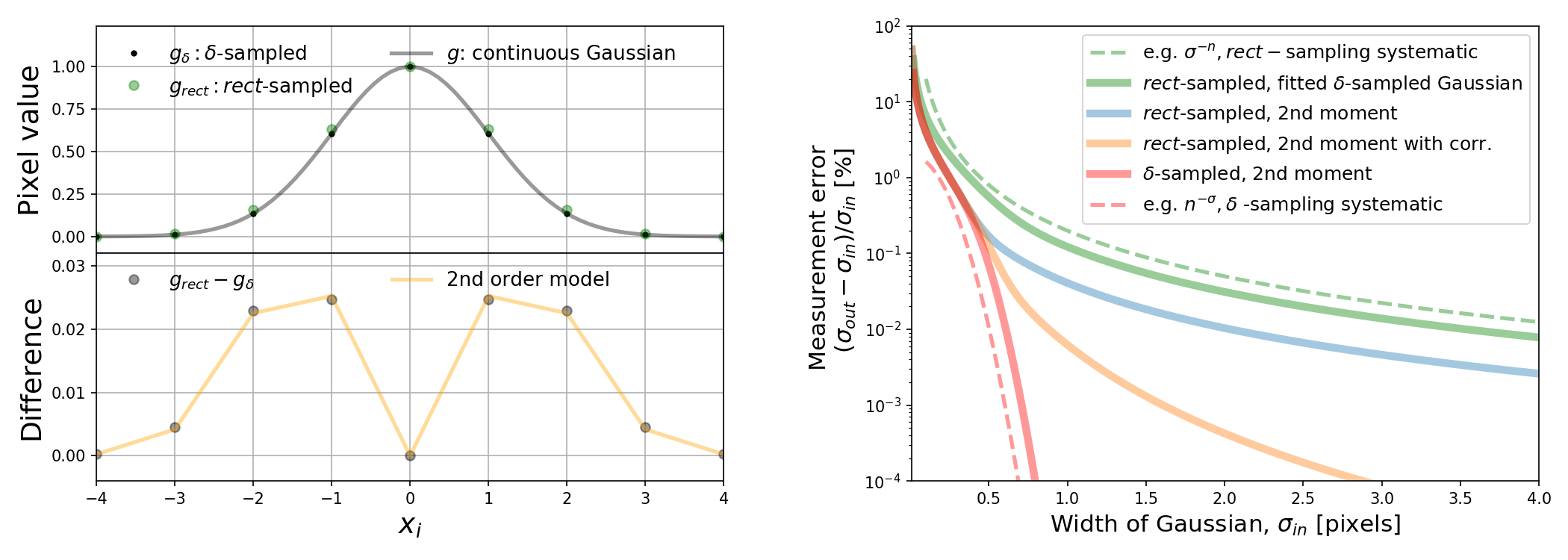}\centering
  \caption{Top left: a Gaussian of width $\sigma=1.0 pix.$ (black line) which has been $\delta$-sampled (black dots) and $rect$-sampled (green points), illustrating the size bias $g_{rect}\geq g_\delta=g(x_i)$. Bottom left: the difference between these images (black points) is compared to the second order Taylor expansion model (yellow line) in Eq. \ref{eq:bias_taylor}. Right: the percentage bias in several estimates of size $\sigma_{out}$ made on $\delta$-sampled and $rect$-sampled images of a Gaussian with input width $\sigma_{in}$. The assumption of a $\delta$-sampling on $rect$-sampled images, for instance in fitting a $\delta$-sampled Gaussian through $\chi^2$-minimization (green) or in typical calculation of second moments (blue), leads to a polynomially-decreasing size bias (green dashed line). Applying the correction model in Eq. \ref{eq:bias_taylor} to $rect$-sampled images can reduce the bias in the second moment calculation at larger input scales (orange). At smaller scales the second moment of $\delta$-sampled images (red) is limited by a pixel spacing systematic which decreases exponentially (red dashed line).}
  \label{fig:bias-rect}
\end{figure}

These 1D images $g_\delta$, $g_{rect}$, and their difference, are shown in the left panels of Figure \ref{fig:bias-rect}. When normalized to the peak pixel value it can be seen that for all pixels $g_{rect}>g_\delta$, implying unweighted moments calculated on the $rect$-sampled image are always biased larger than the input size, i.e. $\mu_{2,rect}>\mu_{2,\delta}>\sigma_{in}$. For instance, the second centralized moment $\mu_{2,rect}$ of a $rect$-sampled Gaussian of width $\sigma_{in}=1$ pixels is biased $0.1\%$ larger than $\sigma_{in}$. Similarly, a systematic bias of $0.5\%$  arises from fitting a $\delta$-sampled model to a noiseless $rect$-sampled image (fitting a $\delta$-sampled model to a $\delta$-sampled image results in zero bias). At input widths greater than a pixel, this bias in second moment of can be decreased using the second order model in Eq. \ref{eq:bias_taylor} (assuming an a-priori knowledge of the centroid and size). As the object size shrinks to scales much smaller than a pixel, the undersampling induces a size bias much greater than a percent, even when second moments are calculated on $\delta$-sampled images. We illustrate these biases in the right panel of Figure \ref{fig:bias-rect} for a range of input Gaussian profiles of width $\sigma_{in}$ and measured by a $\delta$-sampled Gaussian fit, unweighted second moments, and second moments calculated on an image which has had the second order correction (Eq. \ref{eq:bias_taylor}) applied.

In the following sections, we aim to show that pixelization biases such as these have measurable consequences in practical astronomical imaging, from extremely undersampled images encountered in space-based imaging or device characterization, to ground-based survey images which typically aim to critcally-sample the astmospheric point spread function (PSF). We begin in the next section with a description of the algorithms which will be used to compute the moments, including unweighted, Gaussian-weighted, and adaptive moments, as well as a novel set of moments which are correctively weighted to account for bias. In Section \ref{sec:firstmom}, we present the bias in centroid measurement of images and simulations of sub-pixel sized sources which can be found in images with a sub-pixel PSF (e.g. a diffusion-limited PSF). In Section \ref{sec:secondmom} we show that, independent of biases in the centroid, undersampling and pixelization can lead to a systematic bias in size seen in several methods for computing the moment. This bias is shown to be a function of intrinsic object size, and can be partially mitigated using a correctively-weighted moment measurement. In Section \ref{sec:orientation} we show how an uncorrected size-dependent bias in adaptive moments computed on images of elliptical Gaussians can lead to an orientation angle bias pattern. We then demonstrate that such systematic orientation angle bias patterns can be found in modern weak lensing catalogs, breaking the null assumption of isotropic orientation angles. In Section \ref{sec:discuss} we conclude with a discussion.

\section{Algorithms for computing the moment}
\label{sec:algorithms}
Due to its simplicity and generality, the method of moments has long been a popular tool to measure the centroid and shape of objects in images of all types. In astronomy, second moments were first used for weak lensing science to search for the characteristic shear pattern induced by gravitational lensing of distant galaxies\cite{Valdes1983ApJ}. The divergent noise properties of unweighted moments\cite{Kaiser1995ApJ} spurred further development of several variants of weighted moments to measure shapes of objects in a multitude of different surveys of the sky.

Commonly available software such as SExtractor \cite{Bertin1996A&AS} can be used to both automatically detect sources and measure their unweighted and weighted moments in a single pass. To improve the performance and signal to noise of moment measurements, adaptive moments \cite{Bernstein2002AJ,Hirata2003MNRAS} (HSM) were developed and have been used to measure the weak gravitational lensing signal in modern surveys of the sky. In calculating adaptive moments, the weight function is iteratively matched to the measured shape of an image, converging after a given threshold or number of iterations is reached. In the case of a Gaussian weight function, the method is therefore mathematically equivalent to finding the elliptical Gaussian that provides the best least-squares fit to the image. A similar algorithm used in the SDSS Photo Pipeline \cite{Lupton2001ASPC} applies a different metric to stabilize the iteration and test for convergence. We use the \texttt{HSM} adaptive moment algorithm implemented in GalSim \cite{Rowe2015A&C} and a reimplementation of the SDSS algorithms \texttt{SdssShape} \& \texttt{SdssCentroid} available online and in the LSST software stack \cite{Bosch2018PASJ} to measure the bias in computing the second moment on pixelized objects. 


\section{Bias in the centroid} 
\label{sec:firstmom}
Given the discrete locations at which the coordinates and pixels can be sampled, it is intuitively apparent that objects with sizes near the scale of a pixel will not have an accurate centroid estimate using the first moment. This sub-pixel centroid bias encouraged the invention of various mitigation schemes, including weighting, smoothing, and interpolation\cite{Pier2003AJ} which can be found in the \texttt{HSM}, \texttt{SdssShape} and \texttt{SdssCentroid} algorithms. 

\begin{figure}\includegraphics[width=1.0\columnwidth]{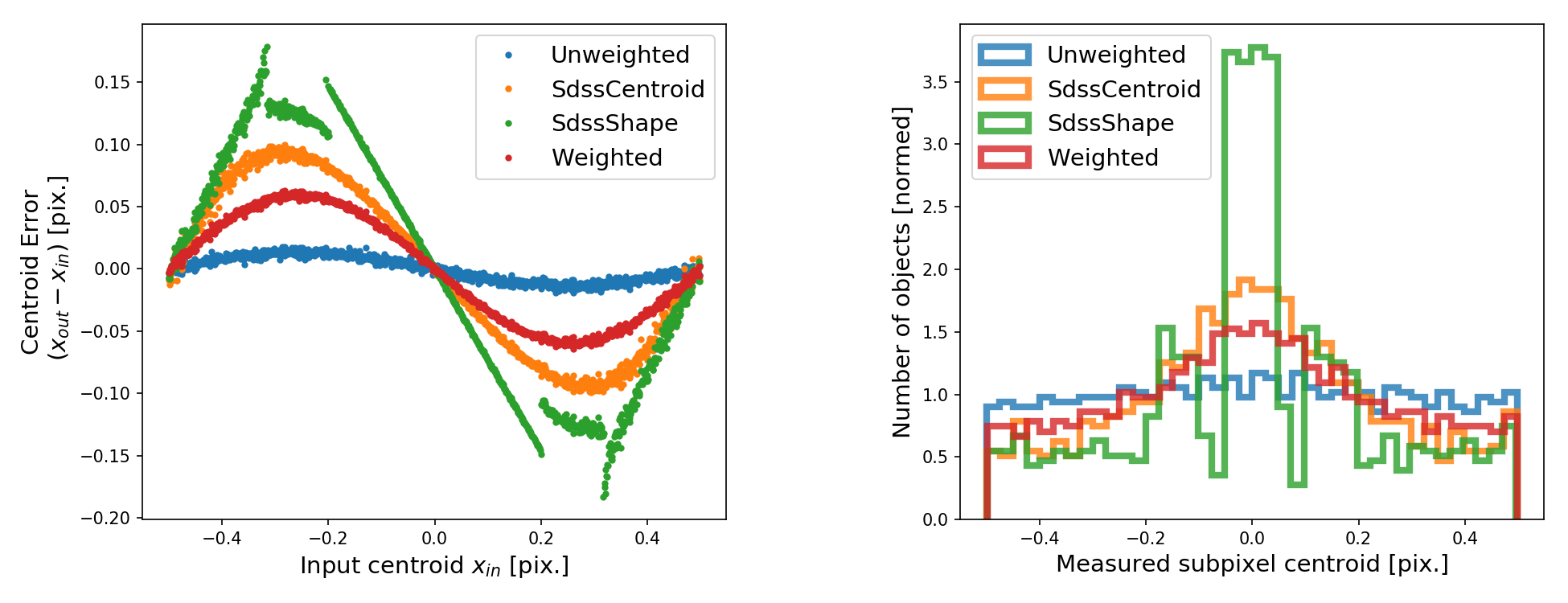}\centering
  \caption[Centroid Bias]{Bias in the first moment computed on dithered and pixelized input Gaussians of width $\sigma_{in}=0.4$ and then measured by a variety of algorithms including an unweighted (blue), weighted with $\sigma_{in} (red)$, and the adaptively-weighted and interpolating methods \texttt{SdssShape} (green) and \texttt{SdssCentroid} (orange). Unweighted moments show the least amount of sub-pixel bias, shown via the difference between the measured and the input $x_{out}-x_{in}$ on the left. This bias translates into an algorithmic `preference' for centroids to be measured near the center of a pixel, as shown in the histogram of sub-pixel centroids on the right.}
  \label{fig:bias-centroid}
\end{figure}

In Figure \ref{fig:bias-centroid}, we show a test of several first moment methods computed on pixelized images of a Gaussian object of width $\sigma=0.4$ pixels, which is roughly the scale of diffusion in an LSST sensor. In the test, the Gaussian image is projected onto the pixel grid at a range of input sub-pixel centroids covering the pixel ($-0.5<x_{in},y_{in}<0.5$ pix.) and then measured with four algorithms: unweighted first moments, weighted using the input Gaussian width, and iterative and interpolative moments via \texttt{SdssShape} and \texttt{SdssCentroid} algorithms.

In this test, it can be seen that the unweighted moments have the smallest bias and that applying a weight function or correction only increases the bias. All methods only produce unbiased estimates when the input centroid is at the middle or the edge of a pixel, and the \texttt{SdssShape} algorithm produces a discontinuous bias when objects are this small, which seems most likely due to the range of validity for the sub-pixel interpolation scheme used in the calculation. The right panel of Figure \ref{fig:bias-centroid} shows the histogram of sub-pixel centroids from these methods, demonstrating a way to diagnose sub-pixel centroid bias in data where the input centroid is unknown. The `preference' for measuring a centroid at the center of a pixel is a common feature which has been observed in measurements made via a variety of methods (including model fitting) made on images of radioactive Iron-55 ($^{55}Fe$) strikes forming diffusion-limited images, as well as sub-pixel pinhole images projected onto CCDs \cite{Christov2019privcomm}. This bias becomes negligible with input Gaussian widths $\sigma$ greater than a pixel, but for extremely undersampled objects this bias remains an issue in all general-purpose algorithms. 


\section{Bias in the size}
\label{sec:secondmom}
\begin{figure}\includegraphics[width=.6\columnwidth]{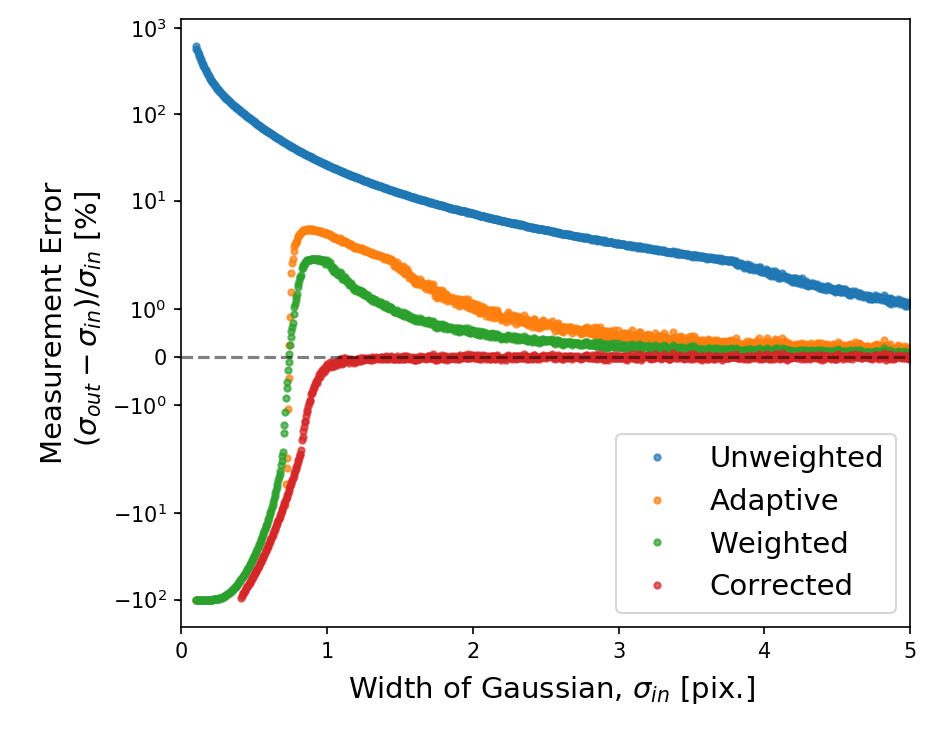}\centering
  \caption[Second moment bias]{Bias in the second moment computed on centered and pixelized input Gaussians of width $\sigma_{in}$ and then measured by a variety of algorithms for computing $\sigma_{out}=\sqrt{m_{22}}$, including an unweighted (blue), adaptively-weighted (orange), weighted with the correct input $\sigma_{in}$ (green), and one weighted using the second-order corrective factor given in Equation \ref{eq:bias_taylor} (red). The correctively-weighted moments illustrate zero bias on scales larger than a pixel, where other methods show a power-law decrease as the Gaussian is well-sampled. On scales smaller than a pixel, all moment methods show a large bias from undersampling.}
  \label{fig:bias-size}
\end{figure}
We illustrate size measurement bias due to pixelization in Figure \ref{fig:bias-size} using images of round pixelized Gaussians of input width $\sigma_{in}$ centered on a pixel, which then have second moments measured by a variety of algorithms including an unweighted, adaptively-weighted, weighted with the correct input $\sigma_{in}$, and one weighted using the second-order corrective factor shown in Figure \ref{fig:bias-rect} and given by Equation \ref{eq:bias_taylor}. The correctively-weighted moments have zero bias on scales larger than a pixel, while other methods show a power-law decrease as the Gaussian becomes better sampled. For these round Gaussians, the \texttt{HSM} and \texttt{SdssShape} adaptive algorithms give identical biases represented by the orange curve in Figure \ref{fig:bias-size}, overestimating the second moment on large scales. This corrective weighting scheme illustrates that the bias can be minimized if a-priori knowledge of the object size and centroid is available or correctly derived from the data.

\section{Orientation angle bias}\label{sec:orientation}
In this section, we demonstrate that a bias in the calculation of size has a measurable effect on the distribution of galaxy orientation angles derived from the second moments. We use the same framework as above, pixelizing elliptical Gaussians with width near the pixel scale $1.05<\sigma_{in}<1.55$, ellipticity $0.2<e_{in}<.6$, and orientation angle unformly distributed $-\pi/2<\theta_{in} <\pi/2$. We measure the second moments $\mu_{2,0}$, $\mu_{0,2}$, and $mu_{1,1}$ using \texttt{HSM} adaptive moments\cite{Hirata2003MNRAS,Rowe2015A&C}, and then compute the orientation angle from this via $\theta = \frac{1}{2} \arctan ( \frac{2\mu_{1,1}}{\mu_{2,0} - \mu_{0,2}})$.

\begin{figure}\includegraphics[width=.9\columnwidth]{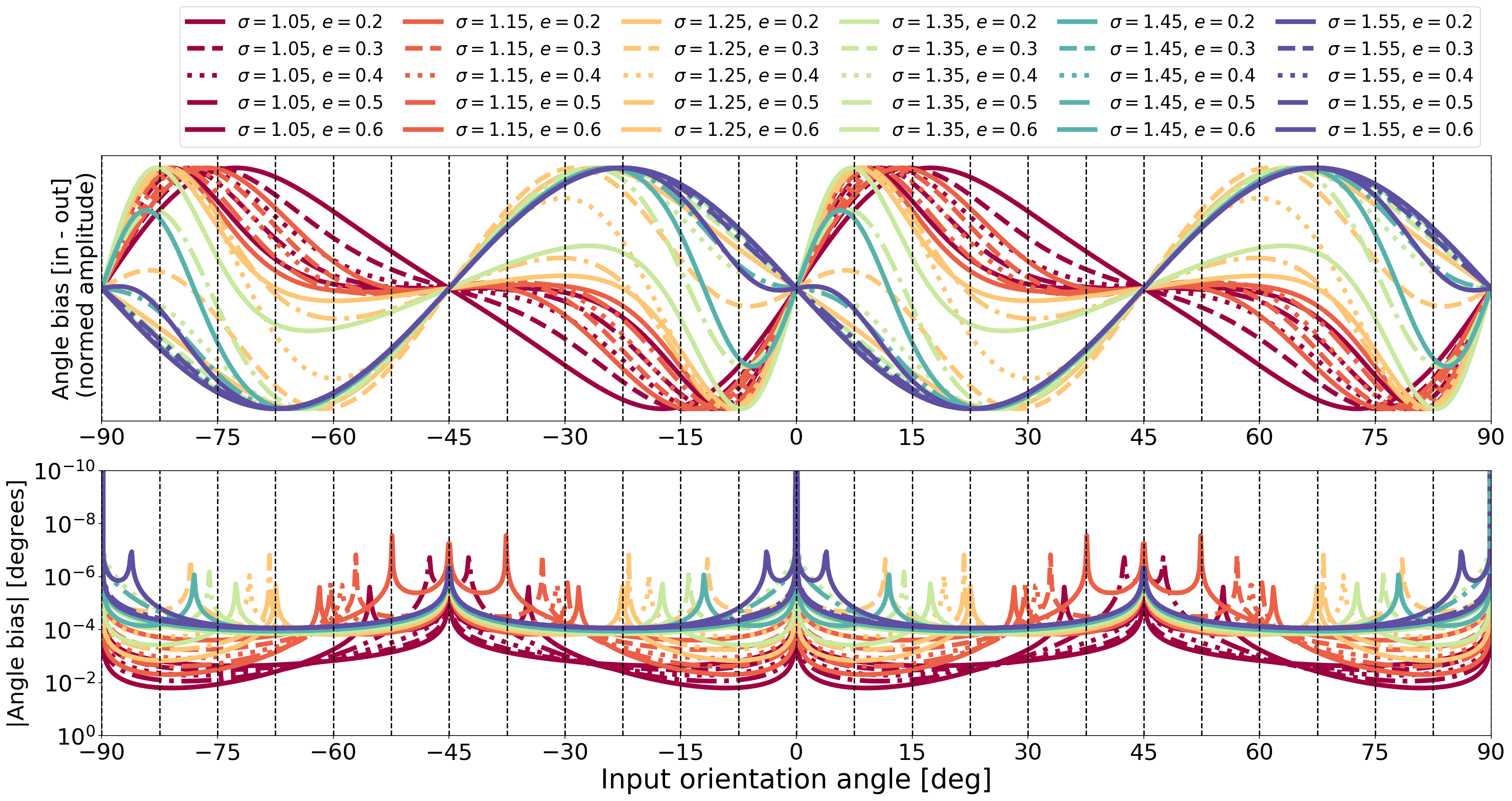}\centering
  \caption[HSM orientation]{Orientation angle bias from computing the \texttt{HSM} adaptive moments\cite{Hirata2003MNRAS,Rowe2015A&C} on pixelized elliptical Gaussian images of a given width $1.05<\sigma<1.55$, ellipticity $0.2<|e|<.6$, and orientation angle unformly distributed between $-\pi/2$ and $\pi/2$. The top figure shows the the difference between the input angle and the one inferred from the moment measurement, normalized to the peak amplitude of the bias for a given set of $[\sigma,e]$. In the bottom figure, the absolute value of these angle bias patterns is shown to illustrate how the bias is minimized at specific angles.}
  \label{fig:orientation-hsm_bias}
\end{figure}

The difference between the input orientation angle and the one computed via second moments of the pixelized image is shown in Figure \ref{fig:orientation-hsm_bias}. It can be seen that all sets of simulated images have some level of bias on the order of $10^{-2}$ to $10^{-4}$ degrees, except at specific angles where the bias vanishes. In all cases, the bias vanishes at $\theta_{in}=\pm\pi/2,\pm\pi/4$, and 0, when the elliptical Gaussian is aligned with the principal pixel axes and therefore the second moments have a minimum bias. However, there are additional angles where this bias is minimized for a given set of $[\sigma,e]$.

A bias which passes through zero represents a transition between an unbiased measurement and one where the bias has opposite signs. This situation is similar to the first moment in Figure \ref{fig:bias-centroid}, where transitions in bias left a signature of peaks and troughs in the histogram of subpixel centroids. Similarly, this angle bias will tend to `push' measurements toward or away from these minimum bias angles. The absolute value of the orientation angle bias shown in the bottom panel of Figure \ref{fig:orientation-hsm_bias} is therefore a sort of mock-histogram of measured orientation angles, where peaks in a histogram are equivalent to minima in a systematic error. If enough galaxies are measured, actual peaks and troughs in the distribution of orientation angles can be observed. This null test of isotropic orientation angles can be used as a probe of systematic error in the shape measurement of objects.

\begin{figure}\includegraphics[width=1.0\columnwidth]{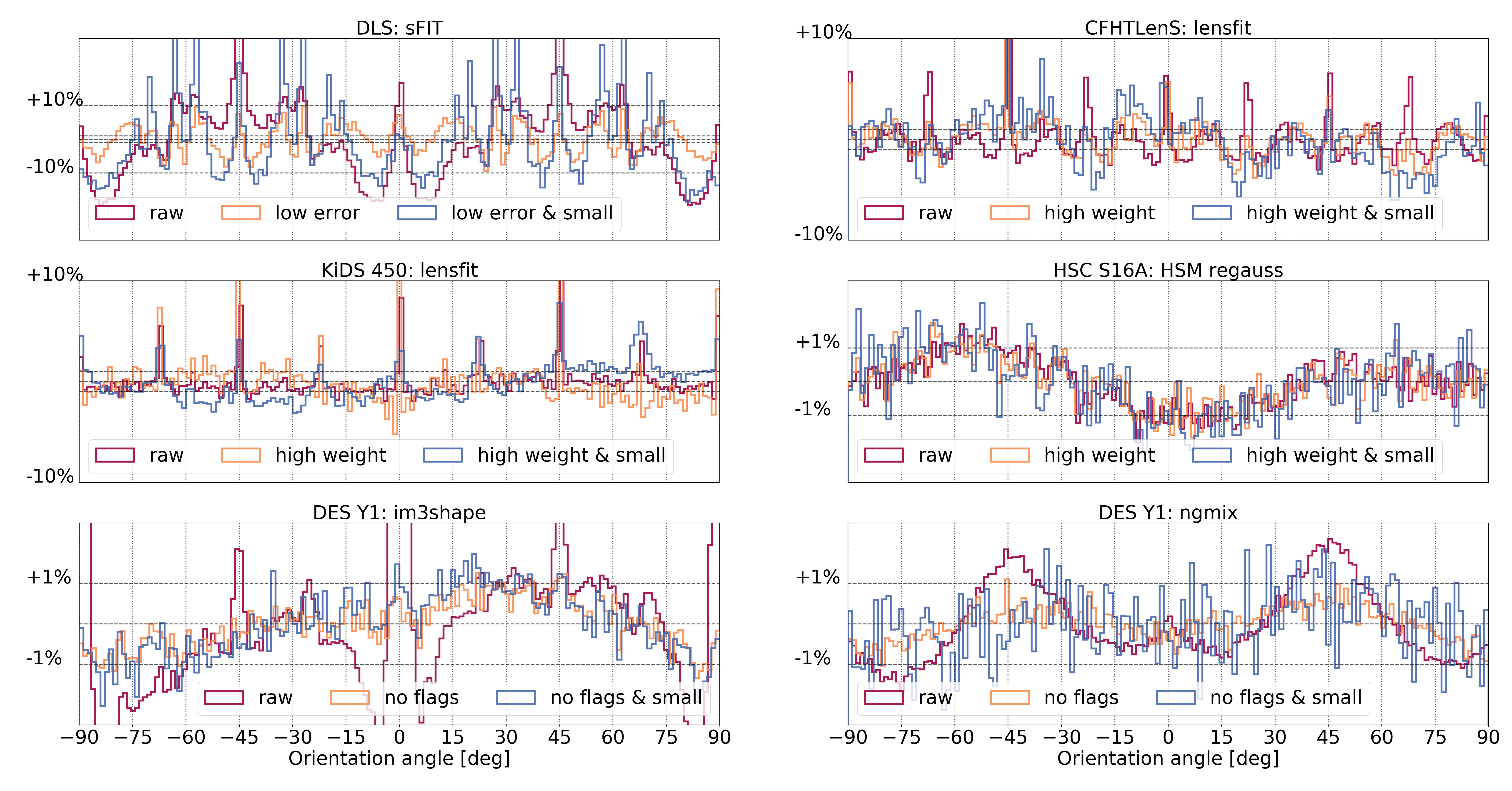}\centering
  \caption[stage three]{The orientation angle distribution of five large sky surveys, where preferential angles can be seen at some level in each. Horizontal dotted lines show the expected average in each bin $\pm 1$ and$10\%$. All survey catalogs have their raw orientation angle distributions shown in red, utilizing all available outputs in the ellipticity catalog. In yellow a more limited sample shown, where either the objects have higher than median weight or lower than median noise, or if they have no flags in the analysis. In blue this sample is further limited to objects which also have a radial size smaller than the median of the catalog.}
  \label{fig:orientation-stage3}
\end{figure}

We perform such a null orientation test using six publicly-available catalogs of uncalibrated galaxy shapes from five surveys. The survey catalogs and the number of objects used, as well as the algorithms which produced them, are listed in the table. These algorithms and their general performance on realistic simulations, can be found in the GREAT3 results \cite{Mandelbaum2015MNRAS} and the papers accompanying the algorithms listed in the table below. Each survey exhibits an orientation bias to some degree, with the amplitude of the effect on the histogram ranging from tens of percent to sub-percent level. As shown in Figure \ref{fig:orientation-stage3}, all surveys show some bias at $\theta=0,\pm\pi/2,\pm\pi/4$, and some at other integer or near-integer fractions of $\pi$. All survey catalogs are first displayed in their raw form, simply using all available valid outputs of the shape measurement algorithm. Additionally, where available, we also display the histograms when the sample is limited to objects without any flags, or where shape measurement weights have greater than the median weight (typically inverse variance) of the catalog. Finally, we also display the pattern of orientation in each survey after a smallness criterion is applied (such that the objects are smaller than the median of the catalog) as well as passing the no-flag/high-weight criterion. Though each of these orientation angle distributions are dependent upon the measurement algorithm and any other cuts made  in the compilation of the catalog, it is observed that the orientation bias generally becomes more prominent when selecting smaller objects inside the same sample.

\begin{tabular}{ |p{6.5cm}|p{3.2cm}|p{1cm} |p{1cm} |p{1cm} |p{1.2cm} |}
 \hline
                         &                     & Seeing & Scale & PSF    & $\sim$ Mag.  \\
 Survey Name &  Algorithm &     [arcsec]   &  [''/pix]  & samp. &  ~~~limit. \\
 \hline
 CFHT Lensing Survey (CFHTLenS)~ \cite{Heymans2012MNRAS,Erben2013MNRAS} & \texttt{lensfit} \cite{Miller2013MNRAS} & 0.7 & 0.187 & 3.7 &  25 \\ 
 Deep Lens Survey (DLS)~ \cite{Wittman2002SPIE} & \texttt{sFIT} \cite{Jee2013ApJ} & 0.8 & 0.26 & 3.1 & 26\\
 Kilo-Degree Survey (KiDS) 450/DR3~ \cite{Kuijken2015MNRAS,Hildebrandt2017MNRAS}  & \texttt{lensfit} \cite{Miller2013MNRAS} & 0.7 & 0.21 & 3.3 & 25 \\
 Hyper Suprime Cam (HSC) DR 1 ~\cite{Mandelbaum2018PASJ} & \texttt{HSM}\cite{Hirata2003MNRAS,Rowe2015A&C}  & 0.6 & 0.168 & 3.6 & 26\\
 Dark Energy Survey (DES) Year 1~ \cite{Zuntz2018MNRAS,Abbot2018ApJS} & \texttt{im3shape} \cite{Zuntz2013MNRAS},\texttt{ngmix} \cite{Sheldon2015ascl.soft} & 1.0 & 0.263 & 3.8 & 23\\
 \hline
\end{tabular}

\section{Discussion and Conclusions}
\label{sec:discuss}
The systematic size biases seen in the introductory 1D model hint at two generic consequences of incorrect assumptions made in measurement of pixelized images. First, the loss of high-frequency information due to finite pixel spacing will bias the estimation of size as seen in the red curve in right panel of Figure \ref{fig:bias-size}. This loss of information due to undersampling introduces a degeneracy between the intrinsic signal and an aliased copy at lower frequency. Second, $rect$-sampled images violate the assumption of a unique correspondence between input parameters and measurements, as the observed image could have been generated by an infinite number of input functions that could provide the same pixel value integral in Eq. \ref{eq:1d_pixel_samp}. These degeneracies between intrinsic parameters and output images leads to a bias in measurements made via model fitting or moment calculations that assume an incorrect pixel sampling or do not account for the information loss at small scales. We propose a simple model to account for the bias in weighted second moment calculations that removes the size dependent bias seen in second moments computed on Gaussians of width $\sigma>1$ pixel or larger.

We demonstrate these pixelization biases using the method of moments computed via various algorithms which are currently in use in astronomy. In Section \ref{sec:firstmom} we show that all forms of calculating the centroid are biased for small sub-pixel sources, and that these biases can be diagnosed via a null test on sub-pixel centroids. Independent of biases in the first moment, the second moment is also biased when the size of objects approach the pixel scale \cite{Bareket79}. Using a second order approximation to the difference between $\delta$ and $rect$-sampled Gaussian images (Equation \ref{eq:bias_taylor}) it is possible to correct for nearly all of the pixelization bias in second moments when a-priori knowledge of the centroid and size of the object is known.

We then demonstrate that an uncorrected size-dependent bias in the second moments can lead to an orientation angle bias, observable as a pattern of peaks and troughs in the distribution of orientation angles computed from biased moments. Similar patterns can be seen in six modern shape catalogs, despite their use of different model-fitting algorithms under a variety of observational, PSF, and pixel-sampling conditions. In fact, similar orientation angle patterns can also be found in generic line fitting algorithm outputs \cite{Desolneux2002IEEE} as well as other studies of orientation angle measurement for cosmic shear. \cite{Whittaker2019arXiv, Whittaker2014MNRAS} Indeed, the simplicity of the noiseless model which resulted in the second moment and orientation angle biases seen in Figures \ref{fig:bias-size} and \ref{fig:orientation-hsm_bias} implies that subtle differences between assumptions and algorithmic implementations can produce the observed orientation bias patterns.

Ultimately, a proof that undersampling is the source of all of these distinctively non-isotropic galaxy angle distributions would require a re-analysis of each survey's data. However, the generic features shared between these orientation bias patterns may be hinting at a common source. For one, all ground-based surveys typically aim to `critically sample' the point spread function (PSF) with the intent of maximizing signal-to-noise. This leaves the majority of distant objects spread across just a handful of pixels. Additionally, in a large surveys of the sky, most galaxies will be at the edge of resolution. Therefore one would expect that less-well sampled imaging surveys might, without a much more careful analysis, lead to more biased orientation angle histograms than better sampled surveys. Whatever the cause of these isotropy-breaking preferential angle patterns, it is clear that further work is needed to achieve unbiased galaxy shape measurement at a sub-percent level.




\acknowledgments      
We gratefully acknowledge Tony Tyson and Craig Lage for their insight. Software used: LSST Data Management Science Pipelines Software \cite{Juric2017ASPC}. This material is based upon work supported in part by the NSF through Cooperative Agreement 1258333 managed by the Association of Universities for Research in Astronomy (AURA), and the DOE under Contract No. DE-AC02-76SF00515 with the SLAC National Accelerator Laboratory. Additional LSST funding comes from private donations, grants to universities, and in-kind support from LSSTC Institutional Members. Financial  support  from  DOE  grant DE-SC0009999 and Heising-Simons Foundation grant 2015-106 are gratefully acknowledged.


\bibliography{article}   
\bibliographystyle{spiebib}   

\end{document}